\documentclass[
    ,final            
  ]
  {aipproc}

\layoutstyle{6x9}

\begin{document}

\title{Conference Summary: \\HI Science in the Next Decade}


\author{Martha P. Haynes}{
  address={530 Space Sciences Building, Cornell University, Ithaca NY 14853}
}

\begin{abstract}
The atomic hydrogen (HI) 21~cm line measures the gas content within and around galaxies, 
traces the dark matter potential and probes volumes and objects that other surveys do not.
Over the next decade, 21~cm line science will exploit new technologies, especially 
focal plane and aperture arrays, and will see the deployment of
Epoch of Reionization/Dark Age detection experiments and
Square Kilometer Array (SKA) precursor instruments.  Several
experiments designed to detect and eventually
to characterize the reionization history of the intergalactic medium
should deliver first results within two--three years time. Although
''precision cosmology'' surveys of HI in galaxies at $z \sim1$ to $3$ require the full
collecting area of the SKA,
a coherent program of HI line science making use of the unique capabilities 
of both the existing facilities and the novel ones demonstrated by the
SKA precursors will teach us how many gas rich galaxies there really are 
and where they reside and will
yield fundamental insight into how galaxies accrete gas, form stars and
interact with their environment.
\end{abstract}

\maketitle


\section{Introduction}
Huge surveys being undertaken at optical and near-IR
wavelengths today are cataloguing and characterizing many 
millions of galaxies, feeding their colors and structural properties
into sophisticated  models that describe their evolution over cosmic
time. In contrast to the wealth of detail available for 
stellar components of galaxies, quantitative measures of the gas 
content are limited to only a few tens of thousands of objects, 
and of the gas distribution within them, to only hundreds. Yet, 
galaxies and the stars within them form out of collapsing gas clouds
and both the amount of gas and its distribution 
within and around galaxies play critical roles in the story of galaxy evolution.

But, how important really is the HI 21~cm line? Sometimes at 
topical meetings such as this one, the attendees are mainly ``the usual 
suspects'' and the elder participants spend most of their time
remembering the ``good old days''. Perhaps to unsettle any tendency 
towards scientific smugness about the long tradition of 21~cm line studies, 
Riccardo Giovanelli, right at the beginning of this conference, reminded
us that HI contributes a ``piffling'' amount even to the overall 
baryon budget in the universe. If that were not enough to
deflate our HI egos, Todd Tripp rightly
pointed out that Ly$\alpha$ absorption lines are a factor of $10^6$ 
times more sensitive to low column density gas than are typical HI maps,
and numerous speakers reminded us that nearly every map we make in HI
pales in spatial detail to the minutely elegant structures gleaned from 
the pretty images made with ``real'' telescopes.
And, as if to render the conference pointless,
Leo Blitz questioned whether HI is really 
relevant to studies of star formation, since stars form out of H$_2$,
not HI. With this self-skepticism as a backdrop to this conference,
speakers and participants
engaged in healthy discussion of the relative importance of HI and
other techniques to cosmology and galaxy studies.
The often-lively discussion was not just of what
we do and don't know, but most importantly, what we {\it might} know
in the next 5-10 years. It
was that focus on peering into the HI crystal ball that
was most appealing: where is HI science headed and how can
we expedite the voyage? In this review, I focus on the outcome of those
discussions, especially as they relate to the the US-based facilities.

\section{The 21st Century: A New Age in HI Spectroscopy}

Two advantages of the radio 21~cm line are its simple physics and
potential detectability over a 
wide redshift range by a single (huge) telescope operating in the 
cm--m$\lambda$ wavelength range.
So, if HI is so ``simple'', why is its application as a tracer of
cosmic evolution so seemingly far behind OIR studies
in terms of redshift numbers and depth? The prime reason is the lack of
facilities: without significant collecting area, HI emission is hard to 
detect at large distances or low column densities. In contrast to
the science potential of small optical telescopes, single small radio 
telescopes cannot contribute to HI studies much beyond 
the Milky Way. For nearly all purposes, HI ``telescopes'' must be big (in terms of
collecting area),
and hence, are expensive both to build and to operate. In practice, instruments rarely
achieve optimal performance in all of the relevant parameters: sensitivity,
sky coverage, angular resolution, field of view, spectral bandwidth and resolution. 
Practical trade-offs compromise some aspect of the science, e.g., collecting area versus
angular resolution, total bandwidth versus spectral resolution.
And, while radio waves are not affected by terrestrial cloud cover, we live
on a planet whose inhabitants create an increasingly noisy radio environment
and in a galaxy which emits a strong synchrotron foreground. 

Most, if not all, of the 
participants of this conference agree that the objective for construction of 
the next major ground-based facility of direct relevance to HI science is the 
``Square Kilometer Array'' (SKA).
With something like a square kilometer of
collecting area, an HI line survey of a billion galaxies from $z = 0$
to 2 or more, as discussed at this meeting by Joseph Lazio, Steve Myers and 
Rogier Windhorst, will be possible. Such a survey, with a peak redshift distribution
at $z \sim 1.5$, will someday allow ``precision cosmology'' measures of
the baryonic acoustic oscillations by an entirely independent spectroscopic
technique as discussed by the Dark Energy Task Force \cite{detf}, while at the same
time tracing the evolution of the galaxy population at epochs where their
baryon content was gas-dominated. Who {\it wouldn't} want to be part of that
great adventure?

The 21st century is bringing to cm--m$\lambda$ astronomy new opportunities
driven principally by advances in computational resources and the development
of array technologies. The latter fall into two quite separate categories:
(1) arrays of beams which occupy the focal plane: feed arrays in which individual
feed horns contribute pixels and focal plane phased arrays which manipulate the
electric field digitally to ``form'' beams; and (2) aperture arrays, which use 
separate antenna elements to synthesize collecting area. Some of the planned
future instruments will incorporate both kinds of array technologies to maximum advantage. 
Attach these devices to digital correlators, add advanced computational capability, 
adopt clever observing strategies and develop powerful signal processing algorithms, 
and you get the SKA, the only trick being that right now, we are still a bit ahead
of the technology wave required to get us our billion galaxies.
Achieving a square kilometer of collecting area
over the entire cm--m$\lambda$ range with a single facility
is an extremely expensive and probably unrealistic proposition.
Rather than discussing what we meant by ``the'' SKA, 
we talked about steps to enable {\it transformational} ``SKA-class'' 
21~cm$\lambda$ HI science, both through the development
of ``SKA precursors'' and through the exploitation of new technologies on the
existing large collecting areas, particularly the Arecibo 305~m antenna 
and the EVLA (Expanded Very Large Array).

Technologically ambitious new instruments including 
ASKAP (the Australian Square Kilometre Array Pathfinder), ATA-350 (the Allen 
Telescope Array expansion), FAST (Five-Hundred-meter Aperture Spherical Telescope),
MeerKat (the Karoo Array Telescope), LOFAR (Low Frequency Array), and
MWA (Murchison Widefield Array), all currently under construction,
promise to validate new technologies and to demonstrate new science capabilities
across the cm--m$\lambda$ range. Each of these facilities will bring
new strengths geared toward particular applications ranging
from low frequency operations to ultrabroad frequency coverage to
wide field imaging. The potential scientific achievement of this suite
of new instruments is quite astounding. Because they are 
truly revolutionary, much technical work remains to be done before their
full capabilities can be realized. In equally impressive ways,
innovative advances in frontends, backends and signal processing capabilities
are making possible gigantic improvements in our ability to explore the extragalactic 
HI universe using the current world-class facilities. 
Where the EVLA and Arecibo demonstrate SKA technologies and techniques 
to explore SKA-class science, they too should be considered SKA precursors and we should
exploit them as such.

The Arecibo 305~m dish remains the world's largest collecting area at
cm--m$\lambda$. As a single
aperture, its strength lies in the detection of weak sources, exploiting its gain
advantage. The advent of the Arecibo L-band Feed Array (ALFA) has thoroughly 
revolutionized not only the kind of surveys underway
at Arecibo but also {\it how} they are being done. John Stocke remarked on my 
career-long predisposition to drift scan observing; like the Sloan Digital Sky
Survey, the Arecibo
Legacy Fast ALFA (ALFALFA) survey\cite{gio05}, demonstrates that 
``keeping it simple'' delivers superior data quality in a highly efficient manner.
An innovation of ALFALFA is the full integration of public multiwavelength 
databases using Virtual Observatory tools into the analysis process, allowing
its catalogs to include both the HI and optical identifications, where such exist.
The comparison of HIPASS (Parkes HI All-Sky Survey\cite{hipass} and ALFALFA\cite{gio07}
\cite{as08}, discussed by Riccardo
Giovanelli, emphasizes the impact of multibeaming on Arecibo's huge collecting area.
The deployment of focal plane phased arrays, when that technology becomes 
available, will energize a similar revolution. 
The development of these devices is vital not only to Arecibo
but also to several of the SKA precursor projects, notably ASKAP, FAST and the 
APERTIF (Aperture Tile in Focus)
upgrade to the Westerbork telescope discussed by Thijs van der Hulst.

Similarly, HI programs are poised to exploit the EVLA in truly exciting ways.
Until now, the VLA correlator could not simultaneously deliver adequate bandwidth 
even to cover the entire velocity range of a typical galaxy cluster and velocity 
resolution was frequently compromised to a point biased against the lowest mass,
narrowest linewidth systems. The new EVLA 
correlator will offer a huge advance for, among other things, HI studies of galaxies 
in and near groups and clusters and deep searches for HI in emission or absorption,
with the potential for commensal observing with all L-band EVLA programs
over the full range from $z =0$ to $0.5$. Its greatly increased capabiliities for
high dynamic range imaging will permit study of HI in the most active,
radio loud systems and searches for very weak absorption lines. An innovative
technical advance to provide prime focus capability would allow the EVLA to 
explore the earlier times $z > 1$, when unlike today, most of the baryons 
in galaxies were in the gaseous component.

As already demonstrated by the on-going ALFA surveys, commensal observing
offers a tremendous increase in the efficiency potential of cm$\lambda$
instruments. The voltage detected by a radio telescope
can be mixed, amplified and sent independently to multiple digital backends 
{\it without loss of signal}, so that the same patch of sky can be examined
for its continuum, time-variable and spectroscopic signatures simultaneously.
As demonstrated by the commensal ALFA surveys being undertaken today at
Arecibo, we should expand the design of programs which can 
exploit this unique feature of radio astronomy. Currently, the galactic HI
TOGS program runs commensally with ALFALFA; by the end of this year, the PALFA 
pulsar survey will turn on a third spectrometer. What a remarkable use 
of a mode that already delivers 99\% open shutter time! 

The SKA is where we are headed, but the path from here to there
must include science productivity along the way. It would
be silly not to work hard toward SKA development; that billion galaxy
survey {\it will be} that revolutionary. But, I suggest that
a square kilometer of collecting area is such a huge leap that we do not
yet understand the evolution of galaxies adequately 
to know how to exploit it fully. So, let's use what we already have
to head down the SKA science path. In the next section, I discuss some considerations
and propose some specific approaches which clearly lead toward the
SKA but keep up a sustained pace of scientific discovery along the way.

\section{HI Surveys: Tracing the Evolution of the Gas}

The new cm--m$\lambda$ instruments currently under construction
are specifically designed to achieve, among other things, the characteristics required 
for the undertaking of major HI surveys. 
Like the SKA, most of them are intended to be used in commensal
mode, particularly with time domain surveys
as described in the RSSKA (Radio Synoptic Square Kilometer Array) 
concept presented at the conference by Steve Myers. 
We can assume that they will deliver superb science, although 
exactly what surveys they will undertake probably remains to be decided based
on final performance specifications and the compromises required to accommodate
commensal survey operations. In addition to these entirely new instruments, the
{\it existing} facilities are also being equipped with new capabilities
which can be tapped to explore a set of {\it different} questions.
I advocate a program of complementary surveys which, in sum, will explore
the complex interrelationships of galaxies and their environments,
stars and gas both within and outside galaxies. As a community,
we should develop a strategy to exploit the few
available facilities in an integrated, coherent science program to address the 
most fundamental questions. The SKA will provide its billion galaxy HI survey, 
but in the meantime, great progress can be made towards understanding the evolution of 
the gaseous constituent of galaxies through coordinated surveys of HI in carefully
selected samples, targeting both individual objects, active, interacting and quiescent
systems, galaxies in and near groups, clusters and voids and in
deep fields. Here are four fundamental questions whose answers will elude us 
unless we view the universe through the HI window:

{\bf How and when did reionization occur?} \hskip 5pt
Assuming we understand anything about astrophysics, before there were
stars in the universe, there was HI, and the most recent constraints set by
WMAP5 \cite{wmap5} imply that reionization took place over an extended
period of half a billion years.
The potential for detection of HI emission from the Dark Ages and throughout
the reionization era shares some analogy with Cosmic Microwave Background science in
the early 1960's, just before the discovering of the 3K radiation.
In the latter case, however, Penzias and Wilson were not deliberately seeking
the CMB whereas, in the EOR case, the theorists are far ahead of the technique.
The current ``first generation''
low frequency experiments designed specifically for this purpose hold great promise
for detection, although we shouldn't underestimate the technical challenges.
In the short term, multiple independent approaches offer the best strategy, until
the first results are in, hopefully with multiple robust -- and consistent -- detections.
The final design of a major EOR {\it facility}, beyond these experiments, must await
their prior results, but it is probably safe to say that throughout the next
decade, we should continue to add collecting area to the first set of instruments, depending
on which ones pan out, what frequencies (redshifts) prove to be of most interest and
what technical solutions offer the most promise.

{\bf How many gas-rich galaxies are there and where do they reside?} \hskip 5pt
Is there a primordial HI mass function (HIMF)? Are there gas-rich ``missing satellites''? 
Do galaxies form in voids? One of the early results of ALFALFA is that
its HI detection rate greatly exceeds that of
pre-survey predictions based on HIMFs derived
from other surveys. We still do not have adequate volume
sampled to conclude whether this is a purely cosmological result
or the product of superior data quality and signal processing software.
However, past widely divergent and sometimes contradictory
inferences on the slope of the low mass slope of the HIMF and 
its possible variation with extragalactic environment
are symptoms of inadequate volume sampling.
ALFALFA will provide cosmologically significant results on
the HIMF overall and should detect several hundred
objects with M$_{HI} < 10^{7.5}$ M$_\odot$. However, it is not
sensitive enough to detect objects like Leo T at a distance greater than 2.5 Mpc.
Remember that the first ``F'' in ALFALFA stands for ``fast''
(its integration time is only $\sim 40$ sec per beam) and that, at the nearer distances, 
solid angle coverage is the best strategy to maximize the surveyed volume \cite{gio05}.
As suggested by Riccardo Giovanelli at this meeting, deeper, wide area Arecibo
surveys should detect significant numbers of low HI mass galaxies throughout 
the Local Supercluster, especially in and near the Virgo
cluster, and in the nearest voids. Among other things,
Arecibo is fortuitously positioned so that Virgo is overhead, allowing a deep
look at the central regions of the Local Supercluster, in and around the Virgo cluster.
Synthesis instruments such as ASKAP, MeerKat, ATA and APERTIF
will provide wide area surveys with the angular 
resolution needed to avoid confusion and to yield HI distributions.
With a systematic, strategic
approach, we {\it could} collaboratively survey the whole sky. I hope we will.

{\bf How and why do some galaxies accrete gas while others don't (anymore)?} \hskip 5pt
From other wavelength tracers, we can infer the star formation
rate (SFR); star formation requires gas. So where {\it did} this gas come from and how did 
it reach a state of collapse? When and why did gas accretion stop? Which baryons end
up in disks and which get blown away? Do interactions drive the evolution in the SFR density? 
What happens to the gas in major mergers? What processes dominate evolution
in different volumes? 
Nearly all work to date on what is happening to galaxies at $z > 1$
lacks quantitative measures of their gas content; our goal should be how
the gas content of galaxies changes during the same epoch $z \sim 1$ to $0$
that the SFR density appears to decline steeply. ALMA will contribute hugely to our 
understanding of the dense gas in galaxies and hence the star formation process, 
but ALMA is {\it not} a survey instrument and molecules need stars to ``light
them up''. What about the gas not actively engaged (yet) in the star formation
process?  As we know from 30 years of studying
HI deficiency, HI is an excellent tracer of the {\it future potential}
for star formation. We need surveys which can provide measures
of the HI mass as input to models of galaxy evolution to complement the
constraints on stellar mass and star formation history contributed by 
colors and equivalent widths from O/IR/UV surveys. We also need a large,
coordinated CO survey so that we can explore possible evolution of the 
molecular-to-atomic gas ratio. 

The galaxies first detected in HI at high redshift will be massive systems.
ALFALFA finds that 99\% of high HI mass galaxies M$_{HI} > 10^{9.5}$ 
can be identified with an optical counterpart; indeed, there is no
theoretical expectation that high mass halos which contain gas
will not form stars. Hence, while HI blind surveys are extremely valuable
for finding low mass, low surface brightness or low luminosity
galaxies which are underrepresented in optically-selected
samples, we can begin the study of the gas content of
galaxies at higher redshift using targeted surveys selected by (known) optical
criteria. Barbara Catinella has shown the viability of detecting HI emission in 
field galaxies at $z \sim$ 0.2--0.3 with modest integration times using
Arecibo's L-band wide receiver. The GALEX-Arecibo-SDSS Survey (GASS) discussed by
David Schiminovich serves as a bridge between ALFALFA and higher
redshift surveys. These Arecibo ventures are just a beginning; it is 
only a matter of telescope time (and, of course, telescope availability).
We do need to be careful in designing surveys to make sure we do
not focus purely on the most gas-rich populations. GASS, for example
will explore galaxies to a limiting mass fraction, M$_{HI}$/M$_{*}$, so
that we can begin to understand why some massive systems have red colors
but still significant gas reservoirs.
And, we should remember that not all early type galaxies are gas-poor:
40\% of the Local Supercluster E/SO's {\it outside Virgo} have HI detected
by ALFALFA, if not coincident with their stellar components, then near them. 
Let's leave our minds open to surprises as to where we find HI.

We should exploit the synthesis instruments, most notably the EVLA, to study 
the distribution of HI in galaxies to probe a a wide range of environments 
especially in and around groups and clusters, in AGN-hosts and interacting systems. 
We need to understand how central supermassive black holes are fed
by accreting gas and why there are relationships between global gas content
and nuclear activity.
We should also exploit the broad bandwidths 
of modern receiver systems to explore the highest redshifts via HI absorption line
studies. First detection and then followup synthesis imaging can allow exploration
of the physical conditions and kinematics within the host galaxies and
in fortuitous circumstances of line-of-sight absorbers. Whatever surveys make 
sensible use of commensal programs should be undertaken; in those cases, our principal
challenge may be to muster the bodies to process and analyze the acquired data.

{\bf How do galaxies convert gas into stars?} \hskip 5pt
What regulates star formation in galaxies of different mass, luminosity and
surface brightness and in different environments? How many stars 
result from feedback processes? Why does the HI gas track so
well the dark matter? 
While HI reservoirs are typically
found at large radii, star formation occurs at smaller radii and scale height 
where the molecular clouds reside.
Yet, as Evan Skillman pointed out, although the HI column
density has little to do with star formation in high mass galaxies,
the opposite is true in dwarfs. In low gas column density disks, we need to 
understand what fraction of stars
is formed as a result of local processes and what of global ones, why some
dwarfs burst while others don't and how stars form in the faintest systems.
Suggesting that the real question is ``How did the
gas arrive at its present state?'', Evan challenged theories of 
star formation to explain both the presence and absence of gas. 
Have all gas-rich ellipticals acquired their gas only recently? Are the
Local Group ``transition dwarfs'' on their way to becoming dSphs?

In combination with observations across the electromagnetic spectrum of
the varied constituents associated with the star formation process,
projects like SINGS, THINGS, LittleTHINGS, ANGST, etc are 
enabling detailed in situ tests of star formation laws. Coordination of
these and future surveys so that they achieve the same spectral resolution, u-v
coverage, sensitivity etc promises the steady accumulation of
an invaluable homogeneous dataset for comparative analysis
of the HI distribution, kinematics and dynamics.
Evan Skillman asked the astute question ``How many such `THINGS' do we need?''
Certainly, many more than contained in the current HI synthesis collection.
As Eric Wilcots has discussed, we need a mix of complete samples
and of classes representative of dwarfs, early/late types, 
group/cluster galaxies, AGNs/starbursts, interacting systems
and isolated objects, a full encyclopedia of the extragalactic HI zoo.

\section{Summary of the Summary}

At this workshop, we discussed what are the
topics in extragalactic science for which HI is relevant and powerful
and what real progress could we make in the near future.
We agreed that the long term objective was more collecting area, but
we also discussed how real {\it transformational} progress could
be made within the next decade. HI studies are {\it vital} 
to the understanding of galaxy 
evolution and provide quantitative clues that {\it no other wavelength}
can. To maximize the potential scientific payoff, we should, as a community,
undertake {\it a set of coordinated large surveys} that are {\it
fully integrated} with efforts to study the stellar and gaseous components
of galaxies via other techniques; some {\it astronomical
socialism} will produce datasets which can be analyzed democratically.
I do not think I am misrepresenting the general sentiment of the
assembled group in claiming that a strategy for HI science over
the next 5--10 years has three key elements:
(1) undertaking of the best science with {\it all} of the SKA ``precursors''
to nail down the role of HI (and dark matter!) in the local universe
and to begin exploration of the evolution of that role over cosmic time;
(2) development of innovative technologies deployed on the existing
and precursor facilities which will lead us toward the SKA; 
(3) engaging the broadest community
to understand and challenge our understanding of how gas fuels
star formation and galaxy evolution.  The third element is the one
least often discussed. Since the HI facilities we can use are
few in number and shared with many other users, the undertaking
of large survey programs must meet ``legacy'' standards but, 
for the health of our science, they must involve a broad cross-section
of the community. Unless we are ready to turn over survey programs
to a few individuals, we must figure out how to engage large
groups in their undertaking without sacrificing quality within fiscal
and practical boundaries.
Our challenge cannot just be ``what to do'', it
also must include ``how to do it''. And ``how'' demands ``strategy'',
``balance'' and ``inclusiveness''. 

Neutral hydrogen may make up a ``piffling'' amount of the total matter
in the universe, but its physical simplicity and the information it conveys
about the gas content, distribution, kinematics and dynamics
opens a powerful window on the evolution of galaxies, of the hierarchical
structures in which they reside and of the universe itself.


\begin{theacknowledgments}
My participation in ALFALFA is supported by NSF grant AST-0607007
and by the Brinson Foundation. The Arecibo Observatory
is part of the National Astronomy and Ionosphere Center which is
operated by Cornell University under a cooperative agreement
with the National Science Foundation. I thank the staff of
the Arecibo Observatory for their ever warm and generous attention
in support of this conference and especially Robert Minchin
for inspiring and organizing it.

\end{theacknowledgments}






\begin{thebibliography}{1}

\bibitem{detf}
Albrecht, A., {\it et al.}, ``Report of the Dark Energy Task Force'',
 \emph{astro-ph/}0609591.

\bibitem{gio05}
  Giovanelli, R. {\it et al.} 2005,
  \emph{AJ} 130, 2598.

\bibitem{hipass}
 Barnes, D.G, {\it et al.},
 2001, \emph{MNRAS}, 322, 486.

\bibitem{gio07}
  Giovanelli, R. {\it et al.} 2007,
  \emph{AJ} 133, 2569.

\bibitem{as08}
  Saintonge, A. {\it et al.} 2008,
  \emph{AJ} 135, 588.

\bibitem{wmap5}
  Dunkley, J. {\it et al.} 2008,
  \emph{ApJS} (submitted); \emph{astro-ph/}0803.0586.
\end{thebibliography}
\end{document}